\documentclass{SCIS2021}
\def \bra#1{\mathinner{\langle{#1|}}}
\def \ket#1{\mathinner{|{#1}\rangle}}

\usepackage{multirow}

\usepackage{appendix}

\usepackage{qcircuit}
\usepackage{setspace}

\usepackage{lipsum}
\makeatletter
\newenvironment{breakablealgorithm}
  {
   \begin{center}
     \refstepcounter{algorithm}
     \hrule height.2pt depth0pt \kern.1pt
     \renewcommand{\caption}[2][\relax]{
       {\raggedright\textbf{\ALG@name~\thealgorithm} ##2\par}%
       \ifx\relax##1\relax 
         \addcontentsline{loa}{algorithm}{\protect\numberline{\thealgorithm}##2}%
       \else 
         \addcontentsline{loa}{algorithm}{\protect\numberline{\thealgorithm}##1}%
       \fi
       \kern.1pt\hrule\kern.1pt
     }
  }{
     \kern1pt\hrule\relax
   \end{center}
  }
\makeatother

\begin{document}
\Year{}
\Month{}
\Vol{}
\No{}
\DOI{}
\ArtNo{}
\ReceiveDate{}
\ReviseDate{}
\AcceptDate{}
\OnlineDate{}

\title{A Variational Quantum Attack for AES-like Symmetric Cryptography}{}

\author[1,2]{ZeGuo Wang}{}
\author[2,1]{ShiJie Wei}{weisj@baqis.ac.cn}
\author[1,2,3,4]{Gui-Lu Long}{gllong@tsinghua.edu.cn}
\author[5]{Lajos Hanzo}{}

\AuthorMark{ZeGuo Wang}

\AuthorCitation{ZeGuo Wang, ShiJie Wei, Gui-Lu Long, et al}


\address[1]{State Key Laboratory of Low-Dimensional Quantum Physics and Department of Physics, \\ Tsinghua University, Beijing {\rm 100084}, China}
\address[2]{Beijing Academy of Quantum Information Sciences, Beijing {\rm 100193}, China}
\address[3]{Beijing National Research Center for Information Science and Technology \\and School of Information Tsinghua University, Beijing {\rm 100084}, China}
\address[4]{Frontier Science Center for Quantum Information, Beijing {\rm 100193}, China}
\address[5]{Department of Electronics and Computer Science (ECS), Southampton {SO17 1BJ}, United Kingdom}

\abstract{We propose a variational quantum attack algorithm (VQAA) for classical AES-like symmetric cryptography, as exemplified the simplified-data encryption standard (S-DES). In the VQAA, the known ciphertext is encoded as the ground state of a Hamiltonian that is constructed through a regular graph, and the ground state can be found using a variational approach. We designed the ansatz and cost function for the S-DES's variational quantum attack. It is surprising that sometimes the VQAA is even faster than Grover's algorithm as demonstrated by our simulation results. The relationships of the entanglement entropy, concurrence and the cost function are investigated, which indicate that entanglement plays a crucial role in the speedup. }


\maketitle

\section{Introduction}
Security of information plays an important role in defense, in the economy and in people's livelihood\cite{FENG D, You X}. At the time of writing typically asymmetric cryptography, such as RSA~\cite{rivest1978method}, is used for transmitting the secret key and symmetric cryptography, such as the Advanced Encryption Standard (AES)~\cite{joan2002design}, is employed for  encrypting data. With the development of quantum computers~\cite{arute2019quantum,zhu2021quantum, CHANG, Kwek}, more and more attention is paid to the security analysis of classical cryptography under quantum attacks.

Shor's algorithm~\cite{shor1999polynomial} is capable of decrypting RSA cryptography in polynomial time~\cite{gidney2021factor}, which seriously threatens the security of asymmetric cryptography. For symmetric cryptography, Grover's algorithm~\cite{grover1996fast,long2001grover,zhu2021robust} can find the key in a set having $N$ entries by only evaluating on the order of $\sqrt{N}$ entries. In Ref~\cite{grassl2016applying,zou2020quantum,wang2021quantum,denisenko2019application}, the efficient quantum implementations of AES and Data Encryption Standard (DES) are proposed by relying on less quantum resources, such as qubits, quantum gates and circuit depths.

At the time of writing we are in the noisy intermediate-scale quantum (NISQ) era~\cite{preskill2018quantum} when quantum computing systems are characterized by low number of qubits, low fidelity and shallow quantum circuits. Under these restrictions, various classical-quantum hybrid algorithms, including the variational quantum algorithm (VQA)~\cite{peruzzo2014variational,yung2014transistor}, and the Quantum Approximate Optimization Algorithm (QAOA)~\cite{farhi2014quantum} have been proposed. These hybrid algorithms have significant advantages in solving combinatorial optimization~\cite{harrigan2021quantum} and Hamiltonian ground state problems~\cite{cervera2021meta}. Briefly, VQA has found applications both in quantum chemistry~\cite{mcardle2020quantum,aspuru2005simulated}, as well as in quantum machine learning~\cite{wei2021quantum,huang2021experimental,beer2020training}, and in quantum finance~\cite{rebentrost2018quantum,egger2020quantum} etc. However, there is a paucity of research on the employment of VQA in classical symmetric cryptography attacks. We fill this knowledge-gap by conceiving a quantum attack scheme based on VQA for AES-like symmetric cryptography. In our design, the parameterized quantum circuit (PQC) operates on the key space, and the cost function is designed according to the known ciphertext. We will show by our simulations that the VQAA on average uses the same order of search-space queries as Grover's algorithm. However in some cases, it is even faster than Grover's algorithm, which is really surprising. We also investigated the relationship between the entanglement entropy, concurrence and the cost function, and it was found that the speedup attained is related to the entropy, which is not unexpected, because the entropy by definition represents the specific degree of surprise upon revealing a particular problem solution/outcome.

The paper is organized as follow. Section 2 briefly reviews the structure of symmetric cryptography and the S-DES technique. Section 3 describes the VQAA process in our work. In this part, the cost function, the ansatzes, and the classical optimization algorithms are designed. Then, the optimization results and the entanglement entropy, as well as the associated concurrence are presented and analyzed. Finally, a summary is provided.

\section{Symmetric cryptography}
The symmetric cryptography, including AES and DES, encrypts and decrypts the data using the same key. For AES~\cite{joan2002design}, there are four operations, AddRoundKey, SubBytes, ShiftRows and MixColumns, to replace and substitute the data. For DES~\cite{Tuchman}, the encryption mainly includes the IP operation, the $f$ function and the SWAP operation. We use the S-DES as an example for characterizing the VQAA for simplicity. The specific process of S-DES is as follows.

\subsection{\label{sec1:level1} S-DES}
The input of S-DES is an 8-bit plaintext and a 10-bit key. The output is an 8-bit ciphertext. 
We can simply express the encryption process as a composite of functions
\begin{equation}
\text{Ciphertext}=\mathrm{IP^{-1}} \circ f_{K_2} \circ \mathrm{SW} \circ f_{K_1} \circ \mathrm{IP} [\text{Plaintext}],
\end{equation}
where $\mathrm{IP}$ is the initial replacement, the function $f_K$ includes the replacement and substitution operations, SW is a swap function and $K_1, K_2$ are the sub-keys in the first and second round of encryption, respectively.
The decryption represents the inversion of encryption in the form of
\begin{equation}
\text{Plaintext}=\mathrm{IP^{-1}} \circ f_{K_{1}} \circ \mathrm{SW} \circ f_{K_{2}} \circ \mathrm{IP}[\text{Ciphertext}].
\end{equation}

\begin{figure}[h]
\center{\includegraphics[width=5cm]{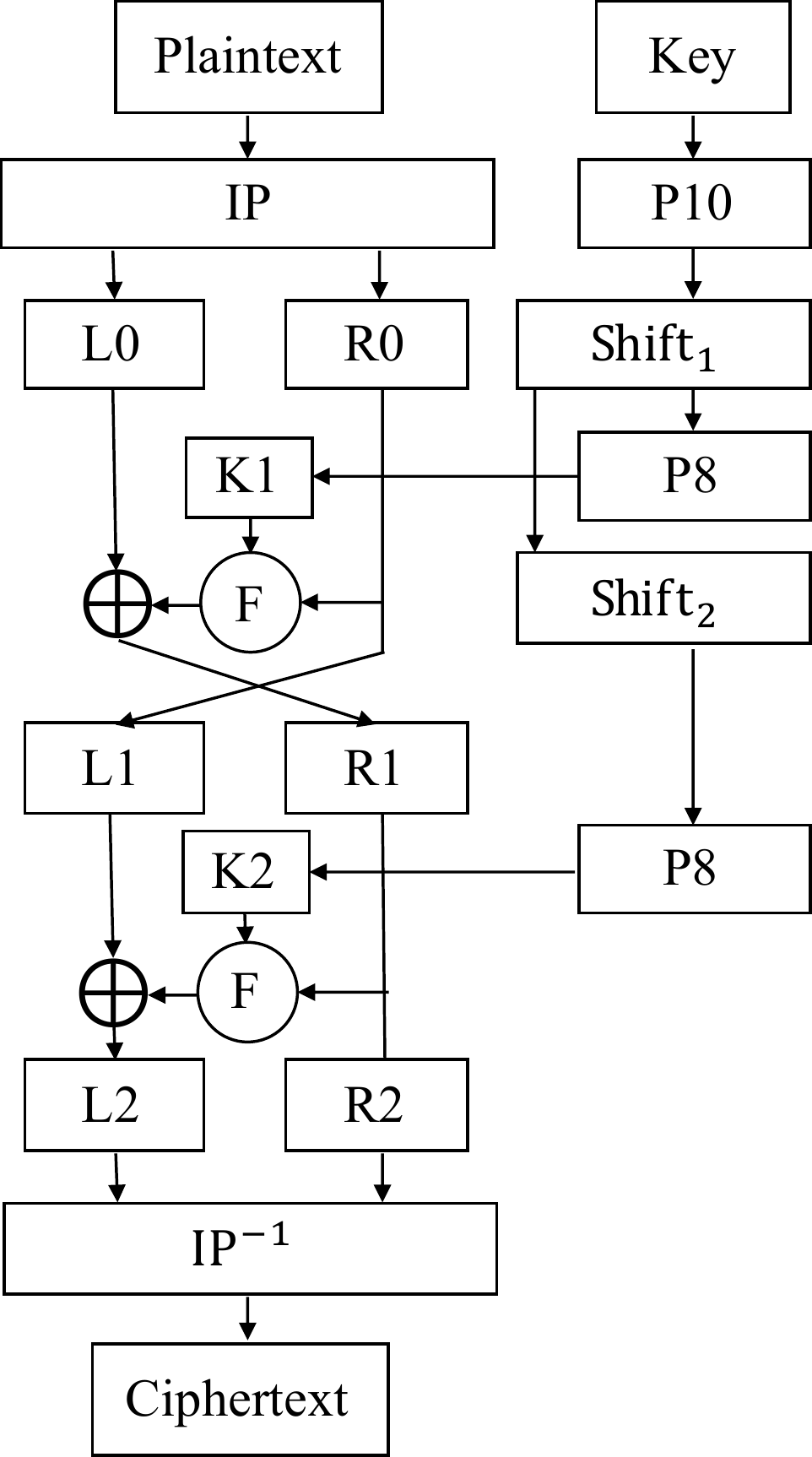}}
\caption{The encryption process of S-DES.}
\label{S-DES}
\end{figure}

The encryption process is shown in Figure~\ref{S-DES} and the associated details are given in~\ref{Sdesdetail}.


\section{\label{Quantum attack}VQAA for symmetric cryptography}
\begin{figure}[h]
\center{\includegraphics[width=15cm]{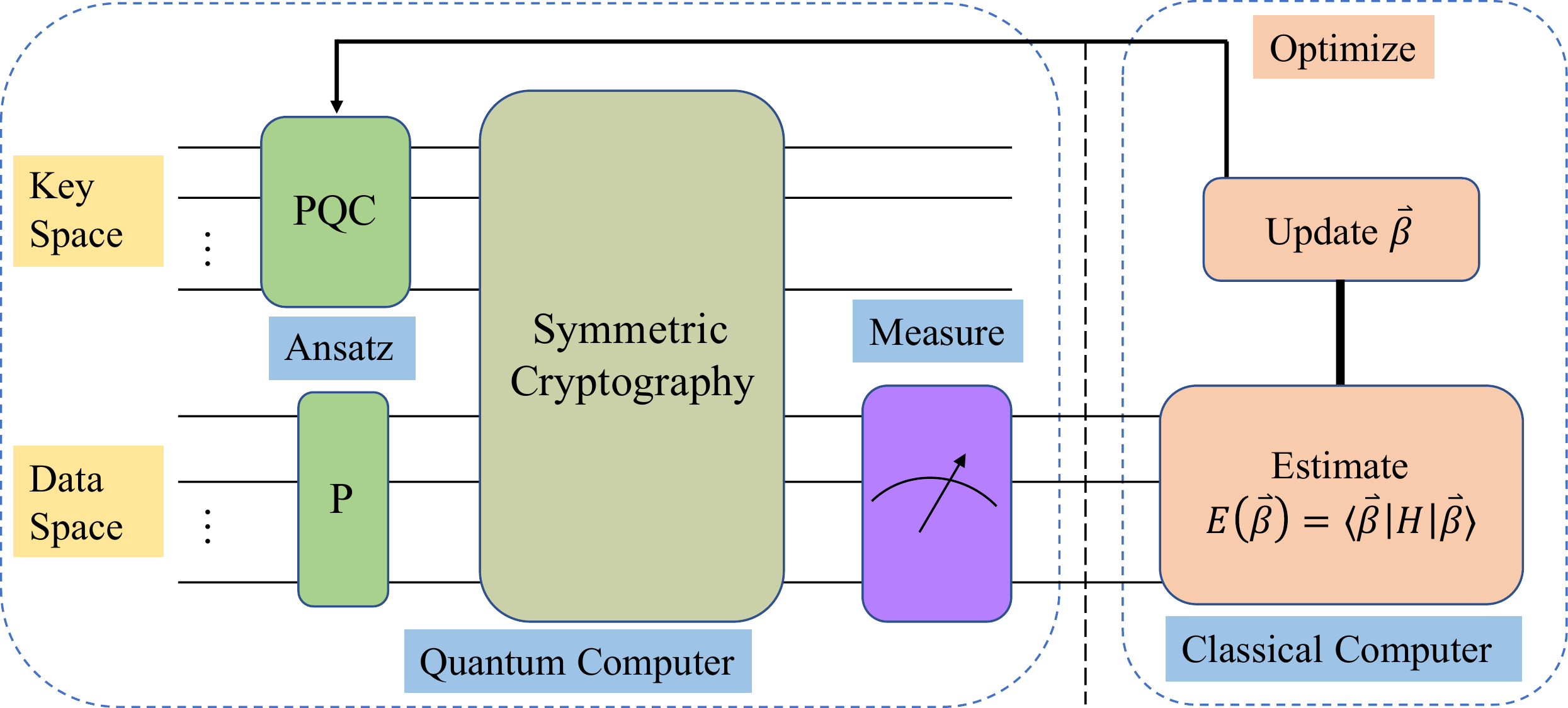}}
\caption{Schematic diagram of the quantum attack on S-DES using VQAs, where $\vec{\beta}$ represents the parameters of PQC.}
\label{process}
\end{figure}

The main idea of our VQAA is shown in Figure~\ref{process}. Based on a pair of known ciphertext and plaintext, the associated Hamiltonian is designed, whose ground state is the ciphertext. The parameterized quantum circuit gives a linear combination of all possible keys. After the symmetric cryptography operations, we have a linear combination of all the ciphertext corresponding to the known plaintext, associated with all possible keys. Then the variational process is started to find the Hamiltonian associated with the lowest energy, which contains the corresponding key.

We now demonstrate how to carry out a quantum attack on the S-DES using the VQAA as an example. The plaintext and ciphertext are represented by 8-bit quantum states. Firstly, we construct the Hamiltonian whose ground state corresponds to the ciphertext. The detailed construction of the Hamiltonian is assumed to be given. Secondly, the key space is encoded into an adjustable quantum state by a parameterized quantum circuit which is also known as ansatz. Next, the output of the parameterized quantum circuit is used as a key to encrypt the known plaintext based on the S-DES and then the superposition of ciphertexts is obtained. Finally, we measure the superposition of ciphertexts and forward the result to the classical optimization algorithm. Using the optimization algorithm, we adjust the input parameters of the parameterized quantum circuit to arrange for the superposition ciphertext state to have a considerable overlap with the known ciphertext. When the result of measurement is the known ciphertext, the key space also collapses to the required key state. In the VQAA of S-DES, the key space and the data space contain 10-qubit and 8-qubit strings respectively, and the 'Symmetric Cryptography' block of Figure~\ref{process} is substituted by the 'S-DES' module whose quantum implementation can be found in Ref~\cite{denisenko2019application}.

\subsection{The construction of cost function}
In order to encode the known ciphertext into a Hamiltonian ground state, we use each bit as a node to construct regular graphs. For an 8-node network, we can construct an $n$-regular ($n$=1,2,...,7) graph. The value of the $i$-th node is denoted by $V(i)$, which is the value of the $i$-th bit. If there is a pair of nodes $i$ and $j$ in the graph that are connected, we add the term $w_{ij} Z_i Z_j$ into the Hamiltonian, where $Z$ is the Pauli-$Z$ operator, $i,j\in \{0,1,...,7\}$. The coefficient $w_{ij}$ is determined by $V(i)$ and $V(j)$, which takes the form
\begin{align}
w_{ij}=&\begin{cases}
 1& \textrm{ if  $V(i) \neq V(j)$},\\ 
 -1& \textrm{ if  $V(i) = V(j)$}.
\end{cases}
\end{align}
Additionally, all of the single-qubit operators $\sum_{i=0}^{7}t_i Z_i$ are added into the Hamiltonian, where $t_i$ is defined as
\begin{align}
t_{i}=&\begin{cases}
 0.5& \textrm{ if  $V(i)=1$},\\ 
 -0.5& \textrm{ if  $V(i)=0$} .
\end{cases}
\end{align}

Then the energy level of these seven Hamiltonians is analyzed. The results are shown in Table~\ref{Hamil}. The 'Ratio' represents the ratio of the energy level differences between the ground state and the first excited state to the total dynamical range of the energy levels.
\begin{table}[h]
\centering
\caption{The energy range of seven Hamiltonians, where 'reg' represents regular. }
\begin{tabular}{cccccccc}
\hline
                                                                                                                                                                                                                                                                                                 & 1-reg  & 2-reg & 3-reg  & 4-reg  & 5-reg  & 6-reg  & 7-reg  \\ \hline
Ground energy                                                          & -8     & -12   & -16    & -20    & -24    & -28    & -32    \\ 
Highest energy                                                         & 4      & 8     & 8      & 9      & 12     & 8      & 4      \\ 
The first excited energy & -6     & -7    & -9     & -12    & -16    & -20    & -24    \\ 
Ratio                                                                   & 0.1667 & 0.2500  & 0.2917 & 0.2759 & 0.2222 & 0.2222 & 0.2222 \\ \hline
\end{tabular}
\label{Hamil}
\end{table}

\begin{figure}[h]
\center{\includegraphics[width=8cm]{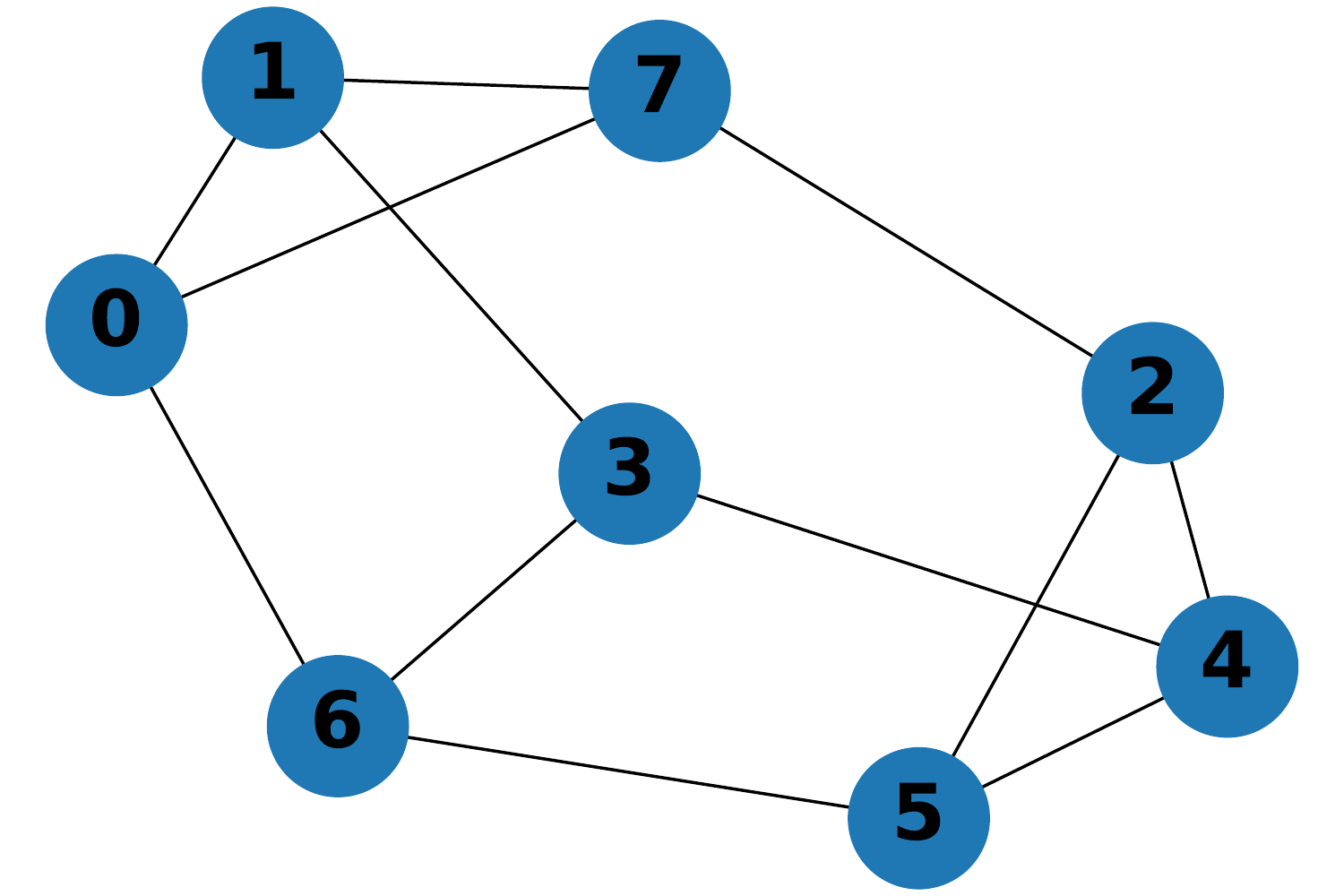}}
\caption{The 3-regular graph used in Hamiltonian construction.}
\label{3regular}
\end{figure}

Instinctively, the higher the ratio, the easier it is to distinguish the global minimum. Because a large ratio implies that the gradient changes rapidly in the vicinity of the minimum value, the optimization path tends to lean more toward the neighborhood of the minimum. The simulations in \ref{regulargraphselect} prove our prediction, when $n=3$, the optimization works best. The 3-regular graph we use is shown in Figure~\ref{3regular}. The number of items in the Hamiltonian is $3 \times 8/2+8=20$, which is a polynomial whose order is determinated by the number of nodes.

The Hamiltonian defined in Figure~\ref{3regular} is
\begin{equation}
\begin{aligned}
H &=w_{01}Z_0Z_1+w_{06}Z_0Z_6+w_{07}Z_0Z_7+w_{13}Z_1Z_3+w_{17}Z_1Z_7  \\
&+w_{24}Z_2Z_4+w_{25}Z_2Z_5+w_{27}Z_2Z_7+w_{34}Z_3Z_4+w_{36}Z_3Z_6 \\
&+w_{45}Z_4Z_5+w_{56}Z_5Z_6 +\sum_{i=0}^{7}t_i Z_i.
\end{aligned}
\end{equation}

As described earlier, $w_{ij}$ and $t_i$ depend on the ciphertext. The cost function $E(\vec{\beta})$ is the expectation of the Hamiltonian,
\begin{equation}
E(\vec{\beta})=\bra{\vec{\beta}} H \ket{\vec{\beta}},
\end{equation}
where $\ket{\vec{\beta}}$ is the superposition of ciphertext state.
\subsection{Ansatz}
\begin{figure}[H]
\centering
\center{\includegraphics[width=13cm]{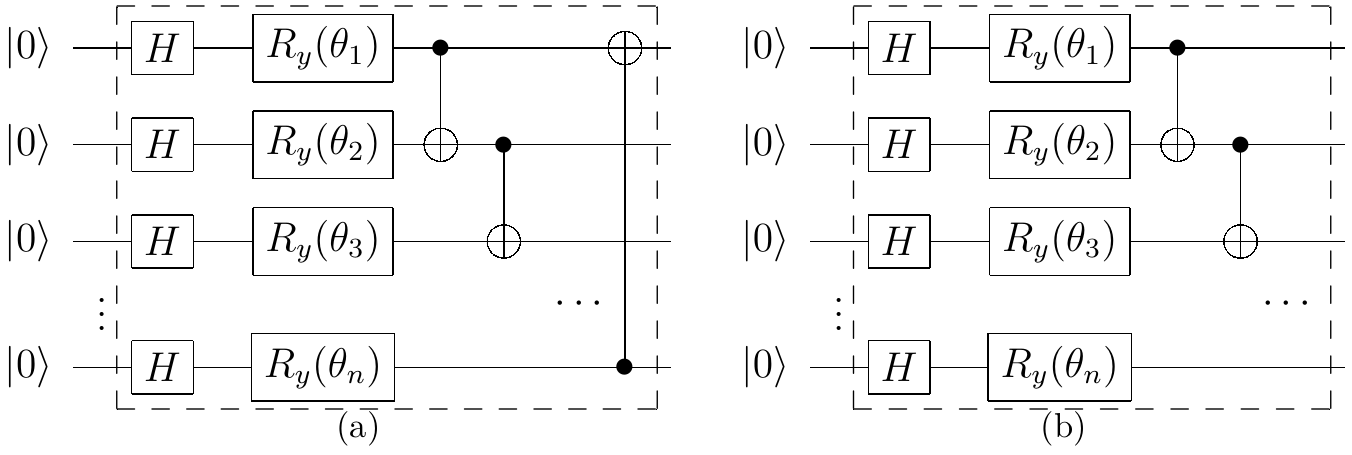}}
\caption{The dashed box indicates a single Y-Cx circuit layer that can be repeated.}
\label{YCx}
\end{figure}

\begin{figure}[H]
\centering
\center{\includegraphics[width=13cm]{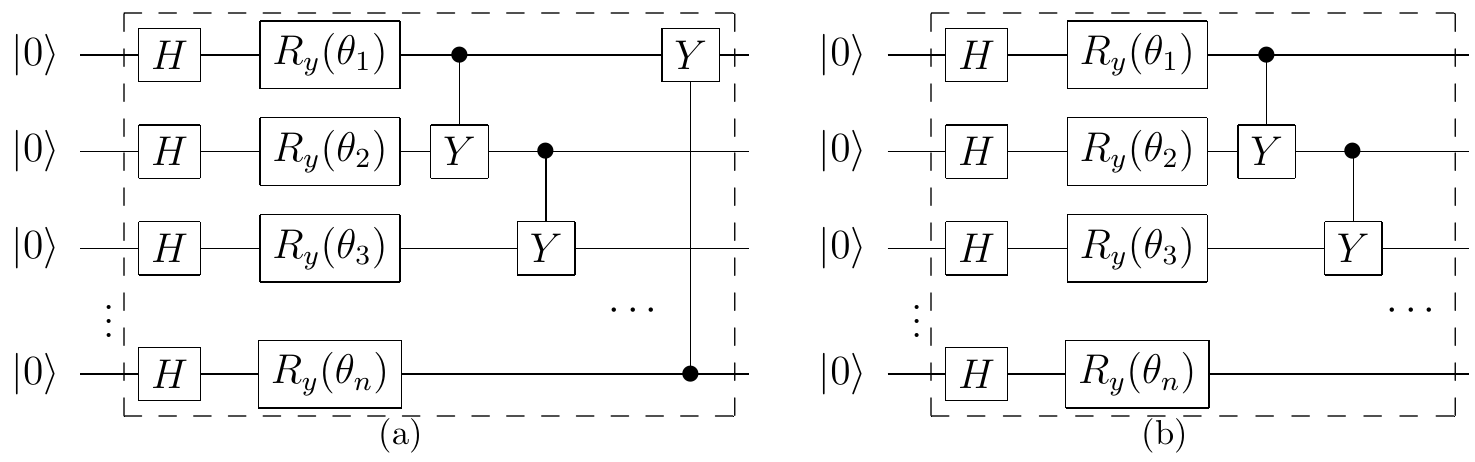}}
\caption{The dashed box indicates a single Y-Cy circuit layer that can be repeated. Gate Y represents a Pauli-Y gate.}
\label{YCy}
\end{figure}

\begin{figure}[H]
\centering
\center{\includegraphics[width=13cm]{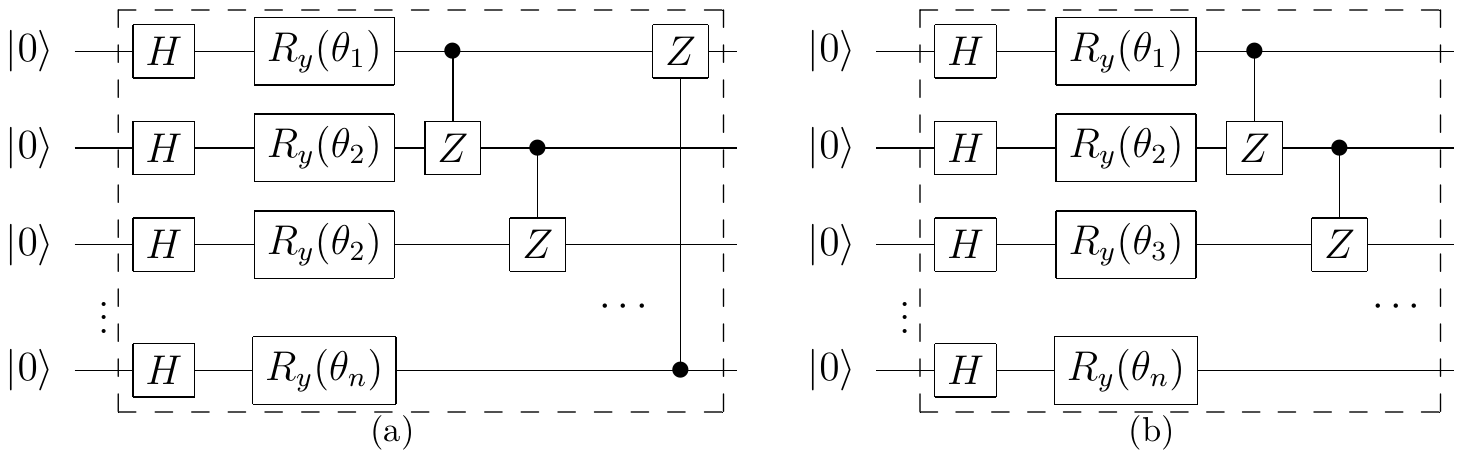}}
\caption{The dashed box indicates a single Y-Cz circuit layer that can be repeated. Gate Z represents a Pauli-Z gate.}
\label{YCz}
\end{figure}

We have chosen six ansatzes in this work. The first two are denoted as the Y-Cx model, which are shown in Figure~\ref{YCx}. The next two are denoted as the Y-Cy model which are shown in Figure~\ref{YCy}, while the last two are denoted as the Y-Cz model which are shown in Figure~\ref{YCz}. The six ansatzes can be divided into two categories, as shown in Figures~\ref{YCx}, \ref{YCy}, \ref{YCz}(a) and Figures~\ref{YCx}, \ref{YCy}, \ref{YCz}(b) respectively. They are denoted as A-ansatz and B-ansatz. Their differences are that there is a controlled $X$, $Y$, $Z$ gate at the right-most edge of Figure~\ref{YCx}, \ref{YCy}, \ref{YCz}(a).

For an $n$-qubit system, a 1-layer ansatz requires $n$ parameters and their circuit depths are $n+2$ for an A-ansatz, and $n+1$ for a B-ansatz. The key space in S-DES contains 10 qubits, so it requires 10 parameters as its input for a 1-layer ansatz and its circuit depth is 12 or 11 for A or B respectively, which is far lower than the depth of S-DES's quantum implementation circuit. The initial state is prepared as the uniform superposition state.

\subsection{Classical optimization algorithms}
We use two methods to optimize the parameters, namely the Gradient Descent method and the Nelder-Mead (N-M) method~\cite{Nelder} whose pseudo-codes and hyper-parameters are given in~\ref{programs}. The cut-off condition is set as $-9$, which is the first excited energy. When the expectation of Hamiltonian is less than $-9$, the superposition cipher state has a large overlap with the known ciphertext (the ground state). When the measurement result is the known ciphertext, the key space collapses to the desired key state. Additionally, we have to set the restart condition for both two optimization algorithms. Explicitly, the VQAA will be restarted when the norm of the gradient is lower than $0.8$ in the Gradient Descent method and when we have $f(x_{N})-f(x_{0}) < 0.15$ for the N-M method. Furthermore, $f(x_{N})$ and $f(x_{0})$ are the maximum and the minimum expectation of the Hamiltonian from the $N+1$ points, respectively. 

\section{The optimization results}
We now characterize the performance of the VQAA for the different combination of ansatzes and optimization algorithms using numerical simulations. The relationships between the cost function and the entanglement entropy, as well as the concurrence are presented.
\subsection{The number of iterations}
\begin{figure}[h]
\centering
\center{\includegraphics[width=15cm]{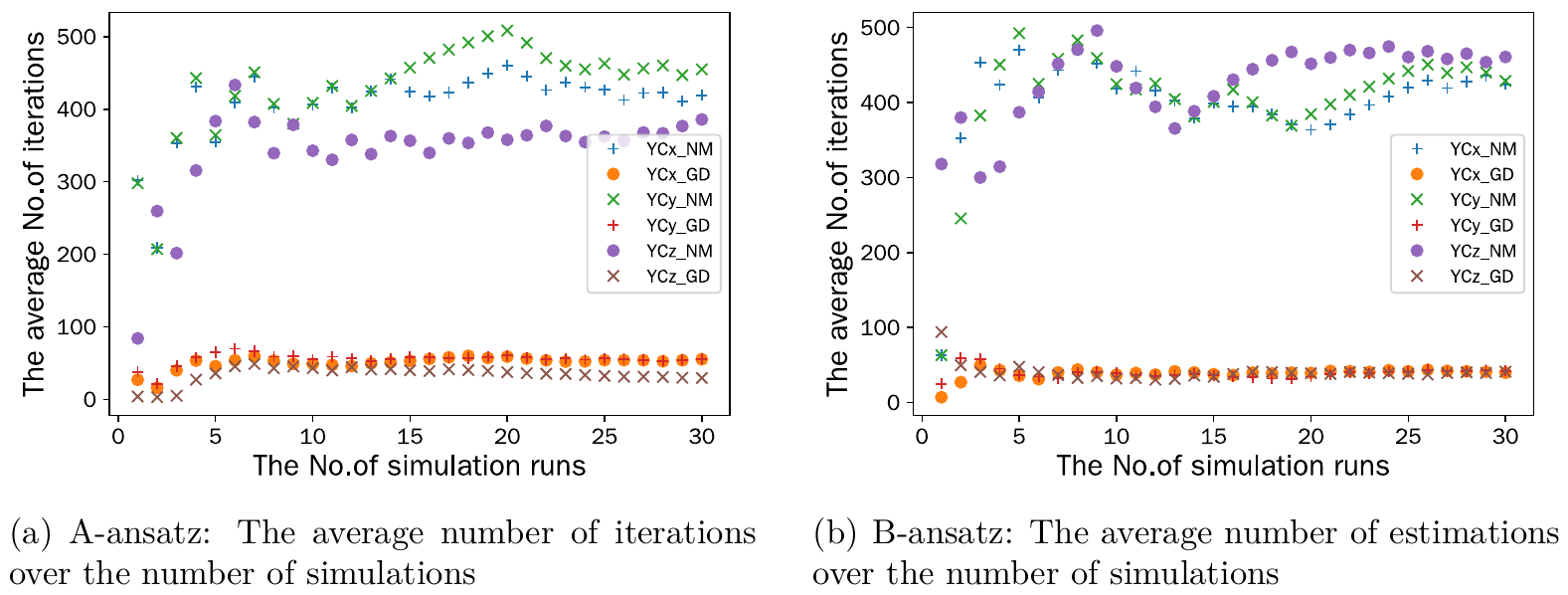}}
\caption{The simulation results}
\label{simulation_times1}
\end{figure}

The hyper-parameters of the classical optimization algorithms are adjusted to the optimal value for the different ansatzes. The initial input parameters are the same in each simulation in which we use different classical optimization methods in order to search for the ground state.
In each simulation, the key and plaintext are chosen randomly, at the same time and the ciphertext is determined. The range of the key is $[0,2^{10}-1]$, and the range of both the plaintext and of the ciphertext are $[0,2^8-1]$. All these values have to be converted into binary strings and then prepared as quantum states. We performed thirty simulations, each with six experiments corresponding to a combination of three ansatzes and two classical optimization algorithms. We terminated the process if we failed to find key after $2^{10}$ measurements. We portray the average number of iterations in Figure~\ref{simulation_times1}(a) and Figure~\ref{simulation_times1}(b) for A-ansatz and B-ansatz, respectively. For the sake of reference, we gave all the number of iterations averaged over 1 to 30 simulations.

\begin{table}[h]
\caption{The number of iterations in the A-ansatz (A) and B-ansatz (B), respectively.}
\label{iterationnumber}
\resizebox{\textwidth}{!}{
\begin{tabular}{cccccccc}
\hline
\multicolumn{2}{c}{\multirow{2}{*}{}}         & \multicolumn{3}{c}{N-M}                                              & \multicolumn{3}{c}{GD}                                               \\  
\multicolumn{2}{c}{}                          & \multicolumn{1}{c}{Maximum} & \multicolumn{1}{c}{Minimum} & Average & \multicolumn{1}{c}{Maximum} & \multicolumn{1}{c}{Minimum} & Average \\ \hline
\multicolumn{1}{c}{\multirow{3}{*}{A}} & Y-Cx & \multicolumn{1}{c}{694}     & \multicolumn{1}{c}{28}      & 419.40  & \multicolumn{1}{c}{94}      & \multicolumn{1}{c}{2}       & 55.27   \\ 
\multicolumn{1}{c}{}                   & Y-Cy & \multicolumn{1}{c}{691}     & \multicolumn{1}{c}{28}      & 454.93  & \multicolumn{1}{c}{94}      & \multicolumn{1}{c}{3}       & 55.67   \\ 
\multicolumn{1}{c}{}                   & Y-Cz & \multicolumn{1}{c}{692}     & \multicolumn{1}{c}{21}      & 385.83  & \multicolumn{1}{c}{94}      & \multicolumn{1}{c}{2}       & 29.50   \\ \hline
\multicolumn{1}{c}{\multirow{3}{*}{B}} & Y-Cx & \multicolumn{1}{c}{705}     & \multicolumn{1}{c}{63}      & 424.53  & \multicolumn{1}{c}{94}      & \multicolumn{1}{c}{4}       & 39.63   \\  
\multicolumn{1}{c}{}                   & Y-Cy & \multicolumn{1}{c}{687}     & \multicolumn{1}{c}{63}      & 428.83  & \multicolumn{1}{c}{94}      & \multicolumn{1}{c}{3}       & 41.93   \\  
\multicolumn{1}{c}{}                   & Y-Cz & \multicolumn{1}{c}{700}     & \multicolumn{1}{c}{19}      & 460.83  & \multicolumn{1}{c}{94}      & \multicolumn{1}{c}{2}       & 41.03   \\ \hline
\end{tabular}
}
\end{table}
In Table~\ref{iterationnumber}, we have given the maximum, the minimum and the average number of iterations required for the six-ansatz and two classical optimization algorithm. It is apparent that the results for those using the Gradient Descent method are much better than those using the N-M method. The Gradient Descent method usually takes 40-50 iterations to obtain the key, whereas the N-M method takes more than 400 iterations. For the six-ansatz scenario, the Y-Cz ansatz's results are better.

\begin{figure}[h]
\centering
\center{\includegraphics[width=15cm]{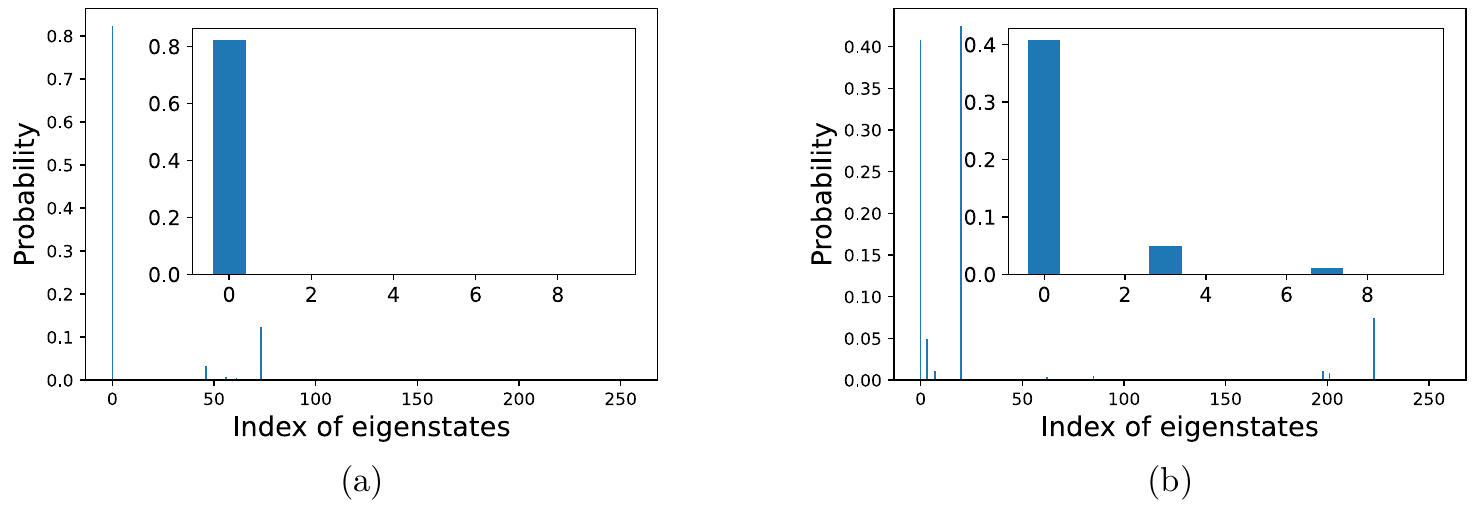}}
\caption{The probability distribution of eigenstates for the Y-Cz(A) ansatz}
\label{Distribution_state}
\end{figure}

When the process is convergent, the occupation probability of the target state is the highest. Figure~\ref{Distribution_state} presents the probability distribution of the eigenstates under the Y-Cz(A) ansatz, when the cost function value is lower than the threshold -9. The $x$-axis represents the eigenstates, which are ordered from the ground state to the highest eigenstate. The $y$-axis represents the corresponding probability.
The probability of ground states in Figure~\ref{Distribution_state} is 0.82 and 0.41, respectively.

\subsection{Convergence vs entanglement entropy/concurrence}
Entanglement entropy and concurrence constitute a pair of popular entanglement metrics. In our work, the relationships between the cost function and the entanglement entropy as well as, concurrence are investigated. For a pure state $\rho_{A B}=|\Psi\rangle\left\langle\left.\Psi\right|_{A B}\right.$, the entanglement entropy is defined as follows,
\begin{equation}
\mathcal{S}\left(\rho_{A}\right)=-\operatorname{Tr}\left[\rho_{A} \log \rho_{A}\right]=-\operatorname{Tr}\left[\rho_{B} \log \rho_{B}\right]=\mathcal{S}\left(\rho_{B}\right),
\end{equation}
while the concurrence is defined as:
\begin{equation}
\mathcal{C}_{A}(\rho)=\sqrt{2\left(1-\operatorname{Tr} \rho_{A}^{2}\right)},
\end{equation}
where $\rho_{A}=\operatorname{Tr}_{B}\left(\rho_{A B}\right)$ and $\rho_{B}=\operatorname{Tr}_{A}\left(\rho_{A B}\right)$ represent the reduced density matrices for each partition. 
\begin{figure}[h]
\centering
\center{\includegraphics[width=15cm]{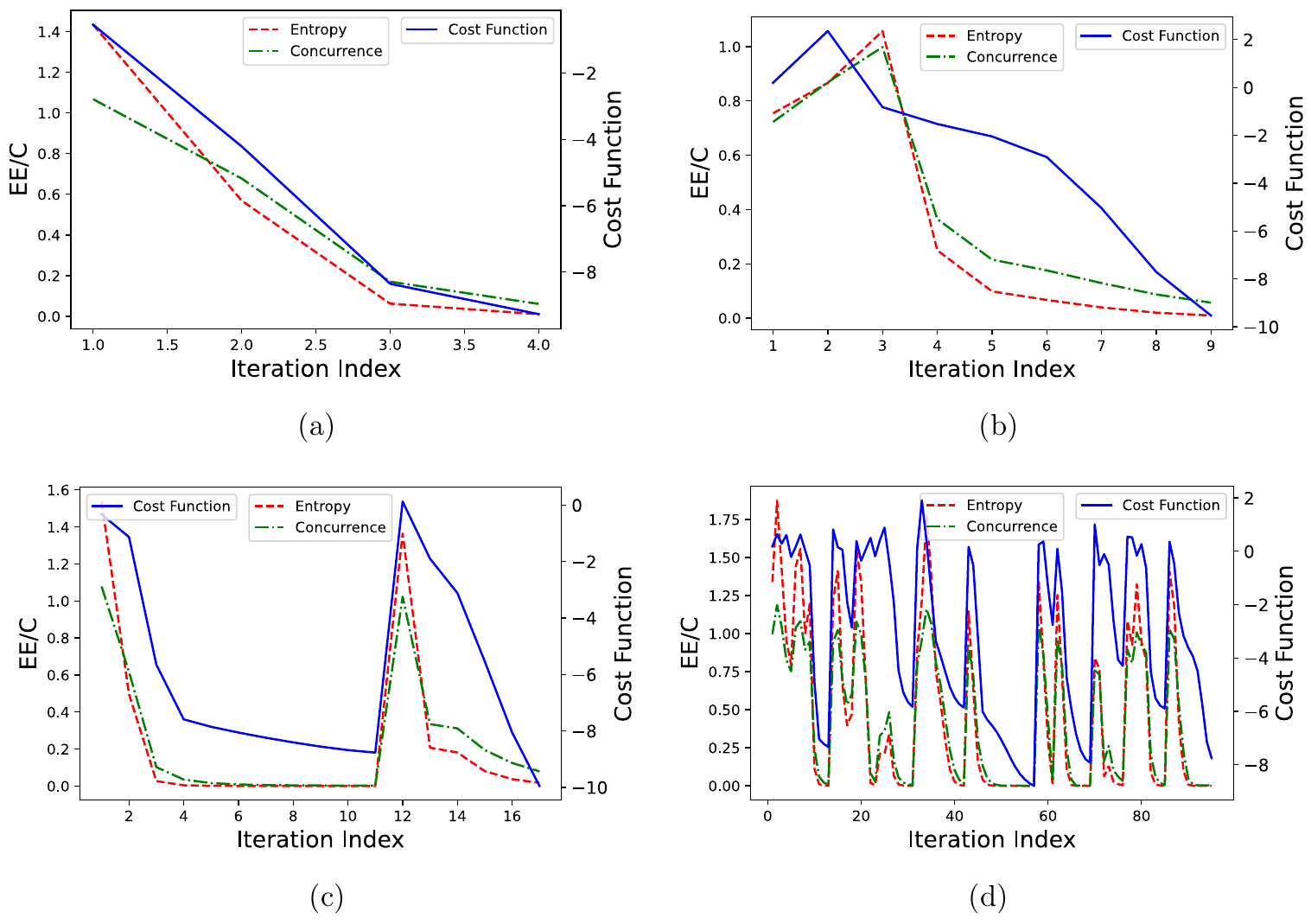}}
\caption{The relationships between the entanglement entropy, concurrence and the cost function, where the y-label $EE$ and $C$ represent the entanglement entropy and concurrence respectively}
\label{EEI}
\end{figure}
We opted for the equal partition, where partition A represents the first 5 qubits and partition B the next 5 qubits, when calculating the entropy/concurrence. 

The results in Figure~\ref{EEI} are generated from the Y-Cz (A-ansatz) and the Gradient Descent method.
There are four scenarios in Figure~\ref{EEI}:
\begin{itemize}
\item[a)]
the process converges monotonically;
\item[b)]
both EE \& C first increase, and then gradually converge towards 0. This case represents a process where the search path is initially close to the target, but there is some initial divergence before convergence; 
\item[c)]
the sudden spike represents a restart. As we recall, we have set up a limit for the gradient to be below at which the iterations eventually are curtailed and then restarted again with a new set of parameters. Nevertheless, the process of searching for the target converges, and becomes successful;
\item[d)]
the process does not converge at all even after several restarts.
\end{itemize} 

As we see from Figure~\ref{EEI}, the entanglement entropy and concurrence behave similarly to the cost function. A strong indication that entanglement is crucial for the success of VQAA.

\section{Summary}
In this work, we proposed a VQAA for orchestrating an attack in symmetric cryptography. We first constructed a cost function whose minimum corresponds to the Hamiltonian's ground state, which is the known ciphertext. The ground state is found by a variational quantum algorithm. Our simulations show that the Gradient Descent method is much faster than the N-M method. The result of Gradient Descent method is comparable to that of Grover's algorithm, and sometimes it is even faster than Grover's algorithm. It will be interesting to investigate further if this trend is also valid when the number of qubits becomes high. If this trend persists when the number of qubits is high, then it is a serious threat to the symmetric cryptography. There are still a lot of open issues, such as the employment of a better classical optimization algorithm, a better ansatz, and better initial parameters. Further studies will continue in the future.

\Acknowledgements{S.W. acknowledges the National Natural Science Foundation of China under Grants No.12005015; We acknowledge the support from the National Natural Science Foundation of China under Grants No.11974205, and No.11774197; the National Key Research and Development Program of China (2017YFA0303700); the Key Research and Development Program of Guangdong province (2018B030325002); Beijing Advanced Innovation Center for Future Chip
(ICFC).}

\Supplements{Appendix A-C. The supporting information is available online at info.scichina.com and link.springer.com. The supporting materials are published as submitted, without typesetting or editing. The responsibility for scientific accuracy and content remains entirely with the authors.
}

\newpage
\appendix
\section{The details of S-DES\label{Sdesdetail}}
\textbf{The generation of sub-key}

Firstly, the initial key can be arranged as ($k_1,k_2,k_3,k_4,k_5,k_6,k_7,k_8,k_9,k_{10}$). The sub-keys $K_1, K_2$ can be generated as 
\begin{equation}
\begin{aligned}
K_{1} &=\mathrm{P} 8[\text { Shift}_1[\mathrm{P} 10[\mathrm{key}]]], \\
K_{2} &=\mathrm{P} 8[\text { Shift}_2[\text { Shift}_1[\mathrm{P} 10[\mathrm{key}]]]].
\end{aligned}
\label{gkey}
\end{equation}
The replacement function $\mathrm{P10}$ is defined as
\begin{equation}
\begin{aligned}
& \mathrm{P} 10\left(k_{1}, k_{2}, k_{3}, k_{4}, k_{5}, k_{6}, k_{7}, k_{8}, k_{9}, k_{10}\right) \\
=&\left(k_{3}, k_{5}, k_{2}, k_{7}, k_{4}, k_{10}, k_{1}, k_{9}, k_{8}, k_{6}\right)
\label{a2}
\end{aligned}
\end{equation}
So the first bit of the output is the third bit of the input, the second bit of the output is the fifth bit of the input, etc. 
Then the first 5 bits and the last 5 bits are shifted to the left by one bit ($\text{shift}_1$), respectively. 
Next, we select 8 bits from the above 10 bits by $\mathrm{P8}$ as the sub-key $K_1$. 
where $\mathrm{P8}$ is expressed as 
\begin{equation}
\begin{array}{cccccccc}
\hline \multicolumn{8}{c} {\text { P8 }} \\
\hline 6 & 3 & 7 & 4 & 8 & 5 & 10 & 9 \\
\hline
\end{array}.
\label{a3}
\end{equation}
In Equation~\ref{a3}, the numbers represent the positions in the 10-bit string of Equation~\ref{a2} and the figures in this section have the same meaning.

Finally, let us return to the 5-bit string pairs produced by the function $\text{Shift}_1$, then shift them to the left by two bits ($\text{Shift}_2$). Turning to the sub-key $K_2$ now, it is obtained by the operation $\mathrm{P8}$ of Equation~\ref{a3}.

\noindent\textbf{The encryption}

For an 8-bit plaintext, the first equation is $\mathrm{IP}$:
\begin{equation}
\begin{array}{cccccccc}
\hline \multicolumn{8}{c} {\text { IP }} \\
\hline 2 & 6 & 3 & 1 & 4 & 8 & 5 & 7 \\
\hline
\end{array}.
\end{equation}
This function retains the 8-bit information of the plaintext but rearranges it. The inversion function of $\mathrm{IP}$ is
\begin{equation}
\begin{array}{cccccccc}
\hline \multicolumn{8}{c} {\mathrm{IP}^{-1}} \\
\hline 4 & 1 & 3 & 5 & 7 & 2 & 8 & 6 \\
\hline
\end{array}.
\end{equation}

The most complex part of S-DES is the function $f_K$, which consists of the replacement and substitution operations. To be specific, $f_K$ is expressed as  
\begin{equation}
f_K(\text{L}, \text{R})=(\mathrm{L} \oplus F(\mathrm{R}, K), \mathrm{R}),
\label{ff}
\end{equation}
where L and R are the left 4 bits and right 4 bits of the 8-bit input, $F$ is a mapping from 4 bits to 4 bits, $K$ is the sub-key, $\oplus$ is XOR operation.
Assuming that the output of the function $\mathrm{IP}$ is 10111101, for some key $K$, we have $F(1101,K)=1110$. Because we have $ 1011 \oplus 1110=0101$, this yields $f_K(10111101)=01011101$.

Now we illustrate the map $F$. Its input is a 4-bit string: ($n_1, n_2, n_3, n_4$). The first operation is its extension
\begin{equation}
\begin{array}{cccccccc}
\hline \multicolumn{8}{c} {\mathrm{E} / \mathrm{P}} \\
\hline 4 & 1 & 2 & 3 & 2 & 3 & 4 & 1 \\
\hline
\end{array}.
\end{equation}
which can be expressed as 
\begin{equation}
\begin{array}{l|ll|l}
n_{4} & n_{1} & n_{2} & n_{3} \\
n_{2} & n_{3} & n_{4} & n_{1}
\end{array}.
\end{equation}
Upon performing XOR with the 8-bit sub-key $K 1=\left(k_{11}, k_{12}, k_{13}, k_{14}, k_{15}, k_{16}, k_{17}, k_{18}\right)$, we arrive at
\begin{equation}
\begin{array}{l|ll|l}
n_{4} \oplus k_{11} & n_{1} \oplus k_{12} & n_{2} \oplus k_{13} & n_{3} \oplus k_{14} \\
n_{2} \oplus k_{15} & n_{3} \oplus k_{16} & n_{4} \oplus k_{17} & n_{1} \oplus k_{18}
\end{array}.
\end{equation}
It can be recorded as
\begin{equation}
\begin{array}{l|ll|l}
p_{0,0} & p_{0,1} & p_{0,2} & p_{0,3} \\
p_{1,0} & p_{1,1} & p_{1,2} & p_{1,3}
\end{array}.
\label{FF}
\end{equation}

The first 4 bits (the first row in Equation.~\eqref{FF}) and the last 4 bits (the second row in Equation.~\eqref{FF}) are imported into the S-box $S_0$ and S-box $S_1$ and the 2-bit output strings, respectively. The pair of S-boxes are as follows:
\begin{equation}
\mathrm{S}_0=
\bordermatrix{%
& 0 & 1 & 2 & 3 \cr
0 & 1 & 0 & 3 & 2 \cr
1 & 3 & 2 & 1 & 0 \cr
2 & 0 & 2 & 1 & 3 \cr
3 & 3 & 1 & 3 & 2
},
\hspace{0.5cm}
\mathrm{S}_1=
\bordermatrix{%
& 0 & 1 & 2 & 3 \cr
0 & 0 & 1 & 2 & 3 \cr
1 & 2 & 0 & 1 & 3 \cr
2 & 3 & 0 & 1 & 0 \cr
3 & 2 & 1 & 0 & 3
}.
\end{equation} 
The first and forth bits of the 4-bit input form a 2-bit binary number, which represents the S-box row, while the second and third bits represent the S-box column. The element determined by the row and column of S-box is the output, which is a 2-bit binary number.

Then the 4-bit output from $S_0$ and $S_1$ is ordered by the operation $\mathrm{P4}$:
\begin{equation}
\begin{array}{cccc}
\hline \multicolumn{4}{c} {\mathrm{P} 4} \\
\hline 2 & 4 & 3 & 1 \\
\hline
\end{array}.
\end{equation}
The output of $\mathrm{P4}$ is the result of the function $F$.

The function $f_K$ just changes the 4 bits on the left, while the Function $\mathrm{SW}$ is an exchange function, which swaps the left 4 bits and right 4 bits of the input, so the input of function $f_K$ in the second round is represented by four different bits. 
~\\
\section{\label{regulargraphselect} The performance of different regular graphs}
In Table~\ref{noitgra}, we show the average number of iterations for different regular graphs where the initial parameters are the same and the other parameters have been set as the proper values. The data is calculated by 15 simulations with the Y-Cz (A) ansatz and the gradient method.
\begin{table}[H]
\centering
\caption{The number of iterations for different regular graphs}
\label{noitgra}
\begin{tabular}{cccccccc}
\hline
                  & 1-reg & 2-reg & 3-reg & 4-reg & 5-reg & 6-reg & 7-reg \\ \hline
Learning rate     & 0.72  & 0.84  & 1.08  & 1.44  & 1.92  & 2.40  & 2.88  \\
Restart condition & 0.53  & 0.62  & 0.80  & 1.07  & 1.42  & 1.78  & 2.13  \\
Maximum           & 94    & 94    & 94    & 94    & 94    & 94    & 94    \\
Minimum           & 4     & 2     & 4     & 2     & 2     & 1     & 14    \\
Average           & 53.27 & 33.07 & 31.13 & 32.73 & 34.87 & 44.93 & 65.60 \\ \hline
\end{tabular}
\end{table}

When the norm of gradient is less than the value of Restart condition, the optimization will be restarted. From the results of Table~\ref{noitgra}, we find the 3-regular graph performs best.

\newpage
\section{\label{programs} Two classical optimization algorithms}
\begin{algorithm}
        \caption{Gradient Descent method}
        \begin{algorithmic}[1] 
        \REQUIRE Initial point $x_0$, the function $f$, the learning rate r, the cut-off condition xerr.
        \STATE times=0
        \STATE Calculate len=length($x_0$)
        \FOR {$ii$ in range(1024)}
        \STATE Let $\text{cost}=f(x_0)$
        \STATE times=times+1
        \IF {costf$<$xerr}
        \STATE break
        \ENDIF
        \STATE Initialize a zero vector Gd, the length is len
        \FOR {$i$ in range(len)}
        \STATE Let $x \leftarrow x_0$
        \STATE Change the $i$-th component of $x$: $x_i=x_i+0.01$
        \STATE $\text{cost}'=f(x)$
        \STATE times=times+1
        \STATE The $i$-th component of Gd: $\text{Gd}_i=(\text{cost}'-\text{cost})/0.01$
        \ENDFOR
        \STATE Generate a random number $r_0$ in range $[0,1]$
        \STATE $x_0=x_0-(\text{r}/|\text{cost}|+\log(\text{times})/\text{times}*r_0)*\text{Gd}$
        \IF {$|\text{Gd}|<0.8$}
        \STATE Initialize a $x_0$ randomly
        \ENDIF
        \ENDFOR
        \STATE 
        \RETURN $x$
        \end{algorithmic}
    \end{algorithm}

* The learning rates in the A-ansatz are set to 0.72, 0.72, 1.08 for the Y-Cx model, Y-Cy model, Y-Cz model respectively; the learning rates in the B-ansatz are set to 0.72, 0.76, 0.94 for the Y-Cx model, Y-Cy model, Y-Cz model respectively.

~\\

\begin{breakablealgorithm}
   \setstretch{}
        \caption{N-M method}
        \begin{algorithmic}[1] 
        \REQUIRE Generating the other $N$ (the dimension of $x$) points ($x_1,...,x_N$) according to the initial point $x_0$. Let the $i$-th component in $x_i$ is $\alpha$ (the amplification factor) larger than the $i$-th component in $x_0$. If the $i$-th component is $0$ in $x_0$, the $i$-th component in $x_i$ is set to $0.8$, the cut-off condition xerr.
        \STATE times=$N$+1
        \WHILE {times$<$1024}
        \STATE Sort and rename these points ($x_i$) in ascending order according to the value of $f(x_i)$, the larger $i$, the larger $f(x_i)$. 
        \IF {$f(x_0) \leq \text{xerr}$}
        \STATE Break
        \ENDIF
        \IF {$f(x_{N})-f(x_{0}) < 0.15$}
        \STATE Restart
        \ENDIF
        \STATE Calculate the average of the first $N$ points $m=\frac{1}{N} \sum_{i=0}^{N-1}x_i$
        \STATE Calculate the reflect point $r=2m-x_N$
        \STATE times=times+1
        \IF {$f(x_1)\leq f(r)<f(x_{N-1})$}
        \STATE $x_N=r$
        \STATE Continue
        \ENDIF
        \IF {$f(r)<f(x_1)$}
        \STATE Calculate the expand point $s=m+2(m-x_N)$
        \STATE times=times+1
        \IF {$f(s)< f(r)$}
        \STATE $x_N=s$
        \STATE Continue
        \ELSE
        \STATE $x_N=r$
        \STATE Continue
        \ENDIF
        \ENDIF
        \IF {$f(x_{N-1})\leq f(r)<f(x_N)$}
        \STATE $c_1=m+(r-m)/2$
        \STATE times=times+1
        \IF {$f(c_1)<f(r)$}
        \STATE $x_N=c_1$
        \STATE Continue
        \ELSE
        \STATE $v_i=x_0+(x_i-x_0)/2$
        \STATE $x_i=v_i$; ($i=1,...,N$)
        \STATE times=times+N
        \STATE Continue
        \ENDIF
        \ENDIF
        \IF {$f(x_N)\leq f(r)$}
        \STATE $c_2=m+(x_N-m)/2$
        \STATE times=times+1
        \IF {$f(c_2)<f(x_N)$}
        \STATE $x_N=c_2$
        \STATE Continue
        \ELSE
        \STATE $v_i=x_0+(x_i-x_0)/2$
        \STATE $x_i=v_i$; ($i=1,...,N$)
        \STATE times=times+N
        \STATE Continue
        \ENDIF
        \ENDIF
        \ENDWHILE
        \STATE \RETURN $x_0$
        \end{algorithmic}
    \end{breakablealgorithm}
~\\
* The amplification factors in the A-ansatz are set to 2.7, 2.7, 2.8 for the Y-Cx model, Y-Cy model, Y-Cz model respectively; the amplification factors in the B-ansatz are set to 2.7, 2.7, 2.7 for the Y-Cx model, Y-Cy model, Y-Cz model respectively.

\end{document}